# Observability of spin precession in the presence of a black-hole remnant kick

Angela Borchers[*], Frank Ohme, Jannik Mielke, and Shrobana Ghosh

*Max-Planck-Institut für Gravitationsphysik*, Albert-Einstein-Institut,
Callinstraße 38, D-30167 Hannover, Germany
and *Leibniz Universität Hannover*, D-30167 Hannover, Germany

Remnants of binary black-hole mergers can gain significant recoil or kick velocities when the binaries are asymmetric. The kick is the consequence of the anisotropic emission of gravitational waves, which may leave a characteristic imprint in the observed signal. So far, only one gravitational-wave event supports a nonzero kick velocity: GW200129_065458. This signal is also the first to show evidence for spin precession. For most other gravitational-wave observations, spin orientations are poorly constrained as this would require large signal-to-noise ratios, unequal mass ratios, or inclined systems. Here we investigate whether the imprint of the kick can help to extract more information about the spins. We perform an injection and recovery study comparing binary black-hole signals with significantly different kick magnitudes, but the same spin magnitudes and spin tilts. To exclude the impact of higher signal harmonics in parameter estimation, we focus on equal-mass binaries that are oriented face-on. This is also motivated by the fact that equal-mass binaries produce the largest kicks and many observed gravitational-wave events are expected to be close to this configuration. We generate signals with IMRPhenomXO4a, which includes mode asymmetries. These asymmetries are the main cause for the kick in precessing binaries. For comparison with an equivalent model without asymmetries, we repeat the same injections with IMRPhenomXPHM. We find that signals with large kicks necessarily include large asymmetries, and these give more structure to the signal, leading to more informative measurements of the spins and mass ratio. Our results also complement previous findings that argued precession in equal-mass, face-on, or face-away binaries is nearly impossible to identify. In contrast, we find that in the presence of a remnant kick, even those signals become more informative and allow determining precession with signal-to-noise ratios observable already by current gravitational-wave detectors.

## I. INTRODUCTION

Mergers of black-hole binaries can produce remnants with significant recoil or kick velocities. This kick is due to asymmetric emission of linear momentum through gravitational waves (GWs). In turn, this process leaves subtle imprints in the GW signal [1–3].

When the initial spins are not aligned with the orbital angular momentum, the orbital plane and the individual spins precess, and the kick magnitudes can reach up to 5000 km/s for specific configurations [4–9]. Nonprecessing binaries, on the other hand, can produce kicks with values up to ∼500 km/s [10,11]. These velocities can become larger than the escape velocities of the remnants' host environments, which impacts the evolution of gravitationally bound environments, such as stellar clusters and galaxies [12–15]. In the case of stellar black-hole binaries, remnant kicks can restrict the possibility of having multiple generation mergers in certain environments (e.g., globular clusters) and can influence the binary black-hole merger rate [16–20].

GW astronomy provides a new way of directly investigating black-hole kicks, complementing the knowledge gained through electromagnetic observations. Several methods have been proposed to extract the kick velocity from GW events, based on the calculation of the linear momentum radiated away by the binary [21,22]. For most of the GW candidates presented by the LIGO-Virgo-KAGRA (LVK) collaborations [23–27], the inferred kick posteriors appear to be uninformative. This is because the kick velocity is strongly dependent on the black-hole spin orientations, which are poorly constrained with current signal-to-noise ratios (SNRs) [27,28]. The uncertainty in our inference of the spin orientations propagates into the kick posterior. Unless the spins are well determined, it is generally difficult to make meaningful kick measurements.

[*]Contact author: angela.borchers.pascual@aei.mpg.de

However, there is a GW candidate with indications of a nonzero kick velocity [29]. GW200129_065458, which we will refer to as GW200129, shows support for a large kick velocity, $v = 1542^{+747}_{-1098}$ km/s. Interestingly, this event also shows evidence for the first clear measurement of spin precession in a GW event [30]. Though it is claimed that the imprint of precession could be mimicked by the presence of an instrumental artifact happening at the LIGO Livingston interferometer [31,32], further glitch-mitigation studies support the evidence of precession in this event [33]. GW200129 is special as it allows to infer meaningful information about the primary spin: the spin tilt angle [30] and the spin azimuthal angle [29]. The authors of Ref. [29] then estimated a kick posterior by using a map from the binary's mass ratio and the spin orientations to the kick velocity. If the intrinsic properties are known, one knows the kick velocity from numerical simulations of similar binaries. Spin precession, on the other hand, is generally challenging to measure as it requires unequal mass ratios, inclined systems, and large SNRs (see, e.g., [34–38]).

In this paper, we study whether the imprint of the kick can help inferring information about the spins. To do so, we investigate spin and kick measurements of simulated binary black-hole signals. We study whether, in the presence of a large kick, one can extract more information about the source parameters from the signal, with a particular focus on the spin measurements. We perform an injection and recovery study where we compare signals with similar spins but significantly different kick magnitudes.

As mentioned above, remnant kicks are caused by anisotropic emission of linear momentum, which is induced by asymmetries happening in the system. This means that there is a preferred direction at merger along which GWs are radiated more strongly. In the case of a remnant that is kicked in the direction of the line of sight towards the observer, one would receive a weaker signal than if the remnant moved away from the observer.

The geometry of the GW emission can be well described by expanding the GW signal into spin-weighted spherical harmonics, also referred to as GW modes,

$$h_+ - ih_\times = \sum_{\ell \geq 2} \sum_{m=-\ell}^{\ell} h_{\ell,m} {}^{-2}Y_{\ell,m}(\theta, \phi). \quad (1)$$

Here, $h_+$ and $h_\times$ are the GW polarizations that are functions of the time or frequency and depend on all source and orientation parameters. In a spherical coordinate system, the dependence on the polar angle $\theta$ and azimuthal angle $\phi$ can be factored out by expanding in the spin-weight $-2$ spherical harmonic functions ${}^{-2}Y_{\ell,m}$. The complex functions $h_{\ell,m}$ are the GW harmonics.

In the case of precessing binaries, the largest kicks are predominantly caused by asymmetries between the $(\ell, m)$ and $(\ell, -m)$ GW harmonics, which are referred to as mode asymmetries. Earlier studies have indicated that neglecting such asymmetries could lead to biased measurements of the source properties [39–41]. However, the exact relation between mode asymmetries and the kick velocity is not well understood. Here, we investigate their relation.

Our study complements Refs. [39–41], as we focus on understanding the impact of the kick. We study equal-mass systems that are oriented face-on to the observer to exclude the impact of higher harmonics on parameter measurements. The emission of equal-mass binary black holes is dominated by the quadrupolar radiation, which has its maximum face-on or face-away to the detectors and its minimum edge-on. Since the emission decreases with the inclination angle, given a false-alarm-rate threshold and assuming the sources are isotropically distributed, face-on (-away) binaries are expected to be observed more regularly than other orientations. Even though those signals might be generated by precessing binaries, measuring precession in these signals is difficult in the absence of higher harmonics (see, e.g., [35,36]). In particular, the authors of Ref. [42] concluded that it is impossible to distinguish between a precessing and nonprecessing binary when equal-mass binaries are oriented face-on or face-away from the detectors.

Besides, the largest kick velocities are produced in equal-mass binaries that undergo spin precession (see, e.g., [5]). In addition, many GW events are expected to be nearly equal mass binaries [27,43], as there are several astrophysical mechanisms that lead to the formation of equal-mass black-hole binaries [44–46]. As equal-mass systems are frequently detected, we might expect some of these signals to include a significant kick.

To generate signals that accurately include the imprint of the kick, it is fundamental to include mode asymmetries in the waveform model. An accurate description of the symmetric waveform is also vital, as inaccuracies in the symmetric waveform can lead to inconsistent kick estimates [47]. Both aspects are key to describe the imprint of the kick in the GW signal. Here, we use the model IMRPhenomXPHM [48–51] and compare it to its enhanced version, IMRPhenomXO4a [52–55], which has a more accurate description of the merger and, contrary to its underlying model, it includes the asymmetry in the dominant mode. This means that IMRPhenomXPHM cannot accurately predict kicks in precessing binaries. Having these models is useful, as we can simulate the same binary with and without a kick by simply using a different waveform model.

Based on our study, we find that the conclusion from [42] is only valid for systems without significant mode asymmetries. In the presence of a significant kick, one can actually measure precession even for equal mass, face-on, and face-away systems. We also find that the two-harmonic formalism presented in [36] does not apply in the case of equal-mass, face-on binaries when using waveform models that include mode asymmetries.

In addition, we study how well we can recover the kick velocity in our injections with IMRPhenomXO4a. The measurability of the kick has already been studied with the models NRSur7dq4 [56] and NRSur7dq4Remnant [57]. These models have the advantage of having a similar accuracy to NR simulations, but they can only be used in a limited region of the parameter space: $M \geq 65 M_\odot$, $q \leq 4$, $|\vec{\chi}_i| \leq 0.8$. Here we use IMRPhenomXO4a, which is calibrated to a larger volume of the parameter space, $q \leq 8$, $\chi_1 \leq 0.8$, and it can be used to generate waveforms for arbitrarily low total masses. This allows us to investigate precessing kicks with a model that can analyze every binary black-hole event observed by the LIGO and Virgo detectors.

## II. METHOD

### A. Parameter estimation

To investigate the inference of binary black-hole parameters in signals with remnant kicks, we perform a Bayesian parameter estimation analysis, the standard method for parameter inference of GW signals. Within the Bayesian framework, the state of knowledge of a specific parameter is described as a probability distribution and is calculated using Bayes' theorem. The posterior probability of a parameter $\theta$ given the data $d$ and a hypothesis $\mathcal{H}$ can be calculated using the following expression:

$$p(\theta|d, \mathcal{H}) = \frac{\pi(\theta|\mathcal{H})\mathcal{L}(d|\theta, \mathcal{H})}{\mathcal{Z}_\mathcal{H}}. \quad (2)$$

Here, $\pi(\theta|H)$ is the prior probability distribution, $\mathcal{L}(d|\theta, H)$ is the likelihood of the data given $\theta$, and $\mathcal{Z}_\mathcal{H}$ is the signal evidence assuming the hypothesis $\mathcal{H}$ is the model for the data. In GW data analysis, one typically assumes stationary Gaussian noise. In such case, one can argue that the likelihood function can be written in terms of the data and a waveform template of the observable signal $h(\theta)$ parametrized by the source parameters $\theta$, and is given by [58,59]

$$\mathcal{L} \propto \exp\left(\langle d|h(\theta)\rangle - \frac{1}{2}\langle h(\theta)|h(\theta)\rangle\right), \quad (3)$$

where we have introduced the inner product between two signals defined as

$$\langle h(\theta)|h(\theta_0)\rangle = 4\Re \int_0^\infty \frac{\tilde{h}(f, \theta)\tilde{h}^*(f, \theta_0)}{S_n(f)} df. \quad (4)$$

Here, $\tilde{h}$ is the signal in Fourier domain, $*$ denotes complex conjugation, and $S_n$ is the noise spectral density of the instrument. In reality, we generate waveform templates of finite length, with a lower frequency limit of $f_{\min} = 20$ Hz and an upper frequency limit of $f_{\max} = 2048$ Hz.

Quasicircular binary black holes are characterized by 15 parameters: eight intrinsic parameters, namely, the individual masses $m_i$ and the spins $\vec{\chi}_i = \vec{S}_i/m_i^2$, and seven extrinsic parameters, the luminosity distance to the source $d_L$, the inclination $\iota$, that is, the angle between the orbital angular momentum and the line of sight, the polarization angle $\psi$, the right ascension $\alpha$ and declination $\delta$ of the source, an arbitrary reference time $t_c$, such as the time of coalescence, and the orbital phase $\phi$ at the reference time $t_c$.

The component masses can also be parametrized by the mass ratio $q = m_1/m_2$, the total mass $M = m_1 + m_2$, or the chirp mass $\mathcal{M} = (m_1 m_2)^{3/5}/(m_1 + m_2)^{1/5}$. Here we use the convention $m_1 \geq m_2$. Two quantities can describe the dominant spin effects of a binary in the radiated GW signal. The effective spin parameter [60–62], $\chi_{\text{eff}}$, is defined as

$$\chi_{\text{eff}} = \frac{m_1\chi_1 + m_2\chi_2}{m_1 + m_2}. \quad (5)$$

$\chi_{\text{eff}}$ quantifies the dominant spin effect in nonprecessing binaries, where the black-hole spins are aligned with the orbital angular momentum. The second parameter is the effective spin-precession parameter [63,64], $\chi_p$, which parametrizes the dominant spin-precession effects in a binary and is defined as

$$\chi_p = \frac{1}{A_1 m_1^2} \max(A_1 S_{1\perp}, A_2 S_{2\perp}), \quad (6)$$

where $A_1 = (2 + 3q/2)$ and $A_2 = (2 + 3/2q)$ are functions of the initial masses and $S_{i_\perp} = |\hat{L} \times (\vec{S}_i \times \hat{L})|$ are the in-plane spin components. Its value is bounded between $0 \leq \chi_p \leq 1$, where larger values represent stronger precession.

### B. Waveform models

In our study we use two different waveform models: IMRPhenomXPHM and IMRPhenomXO4a. They are both frequency-domain inspiral-merger-ringdown models that belong to the phenomenological family, which is based on combining analytical expressions for the early inspiral phase with numerical relativity (NR) data for the merger and ringdown. Hybrid waveforms are produced to later perform phenomenological fits, which are interpolated over the parameter space. Both models include the same set of GW higher harmonics and model precession effects.

IMRPhenomXPHM is designed to model the expected GW signals from precessing binaries. The model includes higher harmonics and is calibrated to an extended set of NR simulations, which gives it a high degree of accuracy. In addition, it incorporates multibanding techniques to accelerate the evaluation of waveforms. Because of its good performance, it has routinely been used by the LVK collaboration to analyze events from GWTC-3.

IMRPhenomXO4a is a new model that adds several improvements to IMRPhenomXPHM: (i) the NR calibration of the $(\ell, m) = (2, \pm 2)$ coprecessing mode, that is, the mode as measured in a noninertial frame that tracks the precession of the orbital plane; (ii) the NR calibration of the precession angles; (iii) the use of an effective ringdown frequency, and (iv) the modeling of the dominant mode asymmetry between the $(\ell, m) = (2, 2)$ and $(\ell, m) = (2, -2)$ coprecessing modes.

The orbit of aligned-spin binaries is symmetric under reflection over the orbital plane, which means that the GW harmonics satisfy a particular symmetry relation,

$$h^*_{\ell,-m} = (-1)^\ell h_{\ell,m}. \quad (7)$$

Such symmetry does not hold in precessing binaries. Instead, there is an asymmetry in the contribution from $+m$ and $-m$ modes to the waveform. On a basis of spin-weighted spherical harmonics, the symmetric ($h^+_{\ell,m}$) and antisymmetric ($h^-_{\ell,m}$) parts of the waveform can be written as

$$h^\pm_{\ell,m}(t) = \frac{h_{\ell,m}(t) \pm (-1)^\ell h^*_{\ell,-m}(t)}{2}. \quad (8)$$

The antisymmetric waveform quantifies the mode asymmetry and is responsible for generating out-of-plane kicks in precessing binaries, as we will show in the following subsection. Out-of-plane kicks can be significantly larger than the in-plane kick produced by the excitation of higher modes of the signal. For this reason, including the antisymmetric part is essential to predict precessing kicks accurately. From the currently existing waveform models, the NRSurrogates and IMRPhenomXO4a are the only models that include this effect.

### C. The remnant kick velocity

To infer the remnant kick velocity of each signal, we compute the radiated momentum flux over the binary evolution, which can be expressed as

$$P_i = -\lim_{r\to\infty} \frac{r^2}{16\pi} \int_{-\infty}^{\infty} dt \oint d\Omega \hat{x}_i(\theta, \varphi) |\dot{h}(t)|^2, \quad (9)$$

where $\hat{x}_i = (\sin\theta\cos\varphi, \sin\theta\sin\varphi, \cos\theta)$ is the unit vector expressed in the spherical harmonic basis and $h$ is the GW strain as defined in Eq. (1). We choose a coordinate system where the orbital plane is in the $x$-$y$ plane, and the $z$ axis is aligned with the orbital angular momentum at a reference frequency. We decompose the GW strain into spin-weighted spherical harmonics, and by integrating over the two-sphere, one can show that the components of the remnant's momentum are given by [65]

$$P_z = -\frac{1}{16\pi} \int_{-\infty}^{\infty} dt \sum_{\ell,m} \dot{h}_{\ell,m}(c_{\ell,m}\dot{h}^*_{\ell,m} + d_{\ell,m}\dot{h}^*_{\ell-1,m} + d_{\ell+1,m}\dot{h}^*_{\ell+1,m}), \quad (10)$$

and

$$P_\perp = -\frac{1}{8\pi} \int_{-\infty}^{\infty} dt \sum_{\ell,m} \dot{h}_{\ell,m}(a_{\ell,m}\dot{h}^*_{\ell,m+1} + b_{\ell,-m}\dot{h}^*_{\ell-1,m+1} - b_{\ell+1,m+1}\dot{h}^*_{\ell+1,m+1}), \quad (11)$$

where the coefficients $a_{\ell,m}$, $b_{\ell,m}$, $c_{\ell,m}$, and $d_{\ell,m}$ read

$$a_{\ell,m} = \frac{\sqrt{(\ell-m)(\ell+m+1)}}{\ell(\ell+1)},$$

$$b_{\ell,m} = \frac{1}{2\ell}\sqrt{\frac{(\ell-2)(\ell+2)(\ell+m)(\ell+m-1)}{(2\ell-1)(2\ell+1)}},$$

$$c_{\ell,m} = \frac{2m}{\ell(\ell+1)},$$

$$d_{\ell,m} = \frac{1}{\ell}\sqrt{\frac{(\ell-2)(\ell+2)(\ell-m)(\ell+m)}{(2\ell-1)(2\ell+1)}}. \quad (12)$$

These are the components parallel and perpendicular to the direction of the orbital angular momentum at the reference frequency. The perpendicular component is a combination of the two planar coordinates, $P_\perp := P_x + iP_y$.

The asymmetries between the positive- and negative-$m$ inertial harmonics are responsible for the net emission of the linear momentum in the $z$ direction, while the interplay of harmonics with different $m$ number is responsible for the in-plane component of the emission of linear momentum. One can compute the kick velocity by dividing the linear momentum by the mass of the remnant black hole.

We generate individual $h_{\ell,m}$ harmonic modes through LALSuite, the LIGO Scientific Collaboration Algorithm Library [66,67], and integrate these to compute the kick velocity using the package SCRI [68–71], where Eqs. (10) and (11) have been implemented.

Alternatively, one could use existing fits for the kick velocity, e.g., NRSur7dq4Remnant, where the remnant kick is computed from the binary's initial masses and spins. However, using these fits on posterior samples obtained with IMRPhenomXPHM or IMRPhenomXO4a can introduce additional systematic errors. We found a disagreement between the kick posteriors estimated from applying NRSur7dq4Remnant on Phenom posterior samples and the kick posterior estimated from integrating the waveform of each sample. Based on this observation, we decided to use the same waveform model for the Bayesian analysis and estimating of the kick posterior. Hence, to infer the kick posterior, we generate a waveform for each posterior sample and compute the remnant's kick

TABLE I. Details of the first set of injections performed. We include the spin magnitudes $a_1$ and $a_2$, the tilt angles $\theta_1$ and $\theta_2$, the angle between the two spin azimuthal angles $\phi_{12}$ (defined at 30 Hz), the effective spin-precession parameter $\chi_p$, the ratio between the maximum amplitude of the antisymmetric and symmetric waveforms with IMRPhenomXO4a, $\max(|h_{2,2}^-|/|h_{2,2}^+|)$, the kick magnitude estimated for these specific spin configurations with the two waveform models, $|v|_{\text{XO4a}}$ and $|v|_{\text{XPHM}}$, and the maximum opening angle $\beta_{\max}$, the maximum value of the angle between the orbital angular momentum and the total angular momentum.

| Injection name | $a_1$ | $a_2$ | $\theta_1$ | $\theta_2$ | $\phi_{12}$ | $\chi_p$ | $\max(|h_{2,2}^-|/|h_{2,2}^+|)$ | $|v|_{\text{XO4a}}$ (km/s) | $|v|_{\text{XPHM}}$ (km/s) | $\beta_{\max}$ (rad) |
|---|---|---|---|---|---|---|---|---|---|---|
| A | 0.748 | 0.748 | 1.007 | 1.007 | 2.789 | 0.63 | 0.39 | 3983 | 0.16 | 0.06 |
| B | 0.748 | 0.748 | 1.007 | 1.007 | 0.145 | 0.63 | 0.03 | 188 | 0.16 | 0.34 |
| C | 0.748 | 0.748 | 1.007 | 1.300 | 3.251 | 0.72 | 0.43 | 3474 | 63 | 0.30 |
| D | 0.748 | 0.748 | 1.007 | 1.300 | 0.053 | 0.72 | 0.03 | 180 | 33 | 0.37 |
| E | 0.748 | 0.748 | 1.007 | 0.973 | 3.254 | 0.63 | 0.39 | 203 | 0.19 | 0.03 |

velocity by integrating these waveforms as expressed in Eqs. (10) and (11).

The injected kick magnitudes are included in Table I. As expected from the lack or inclusion of mode asymmetries, respectively, IMRPhenomXPHM and IMRPhenomXO4a have different kick estimates for the same configurations. We infer the kick prior by applying the same procedure to samples of the source properties (e.g., the individual masses and spins) drawn from their respective priors, which we define in Sec. II E. For consistency, we generate waveforms with the same model used for the parameter estimation analysis. As the mapping from binary properties to the final kick differs between waveform models, we find that the kick prior distributions are also different for each model. The prior computed with IMRPhenomXPHM only reaches kick values of ∼400 km/s, while the prior of IMRPhenomXO4a takes values up to ∼4000 km/s.

In the following, we will use the wording *large kick* to refer to $v \geq 1000$ km/s. These are kicks that can only be produced in precessing binaries. Similarly, we use the term *small kick* for $v \leq 500$ km/s, the range of values produced in nonprecessing and some precessing binaries. We perform two injections with large kicks and two with low kicks. As we fix to equal-mass, face-on binaries, with kick magnitudes in the lower or upper end of the range of possible kick velocity values, these are not random points in the parameter space, but selected points that are representatives of the possible kick magnitudes and their effect on parameter estimation. Any other equal-mass binary can be understood by the cases studied here.

### D. Interplay between mode asymmetries and the kick velocity

We now discuss how the amplitude of the dominant antisymmetric contribution, $|h_{2,2}^-|$, is related to the kick velocity. We compute both quantities for a set of equal-mass binaries with different spin configurations using IMRPhenomXO4a. We keep one spin vector and the magnitude of the second spin fixed while sampling over the second spin tilt, $\theta_2 \in [0, \pi]$, and azimuthal angle, $\phi_2 \in [0, 2\pi]$. We generate the antisymmetric waveform with the function SimIMRPhenomX_PNR_GenerateAntisymmetricWaveform available in LALSuite, which we inverse Fourier transform to the time domain. For a more intuitive understanding of the antisymmetric amplitude, we use the ratio between the maximum amplitudes of the antisymmetric and symmetric mode contributions.

In Fig. 1 we show the relation between the kick velocity and the ratio between the antisymmetric and symmetric waveforms as a function of the secondary spin azimuthal angle, as indicated by the colormap. Here, we fixed the primary spin to $\vec{\chi}_1 = (0.2, -0.6, 0.4)$, and the secondary spin magnitude to $a_2 = 0.748$. What is interesting for our study is that small asymmetries are only compatible with small kicks, while large asymmetries do not necessarily mean large kicks. Signals with the same antisymmetric amplitude might have significantly different kick velocities, depending on the relative phase between the symmetric and antisymmetric waveforms. An intuitive way of understanding the impact of this phase difference on the kick is described in [72]. Yet, large kicks are only generated by large asymmetries. We find that these statements remain valid when changing the binary's mass ratio. Further investigations on the relation between kicks and mode asymmetries will be presented in a forthcoming publication.

### E. Injections

We perform an injection and recovery study with equal-mass configurations. To understand whether the presence of a kick can have any impact on the estimation of source parameters, we compare binary black-hole signals with significantly different kick velocities but with spin orientations that are as similar as possible. In particular, we want these binaries to have the same precession parameter $\chi_p$, which is defined in terms of the spin magnitudes and the spin tilts, and is independent of the spin azimuthal angle.

As shown in Fig. 1, the kick velocity is highly sensitive to the spin orientations, in particular, to the spin azimuthal angles. By modifying the angle between the two spin azimuthal angles, $\phi_{12}$, and keeping the remaining parameters fixed, we can find binaries with significantly different kick velocities, but with the same $\chi_p$ value. Therefore, we compare signals from binaries with the same spin magnitudes and spin tilts, but with different spin azimuthal angles.

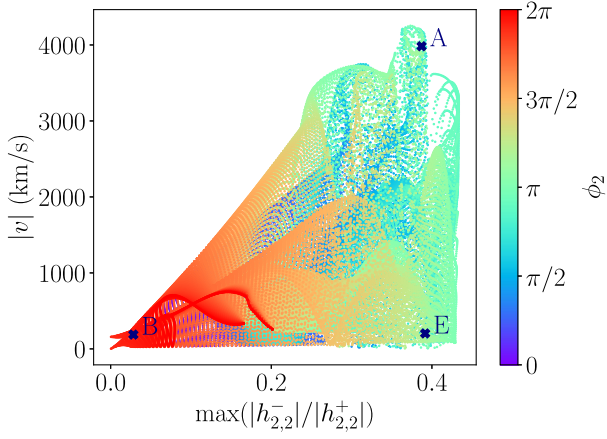

FIG. 1. Relation between the kick magnitude and the antisymmetric waveform with IMRPhenomXO4a. The colormap indicates the dependency of these two quantities on the secondary spin azimuthal angle, $\phi_2$. To produce this figure, we fixed the mass ratio $q = 1$, the primary spin vector $\vec{\chi}_1 = (0.2, -0.6, 0.4)$, the secondary spin magnitude $a_2 = 0.748$, and sampled over the secondary spin tilt and azimuthal angles. The dependency on the secondary azimuthal angle $\phi_2$ can also be seen as a change in $\phi_{12}$. The crosses indicate the parameters of three different binaries used in the Bayesian analysis: injection A (top cross), injection B (lower left cross), and injection E (lower right cross).

To investigate the impact of the kick, we compare the parameter inference of a signal with a large kick with that of an equivalent signal with a small kick. As listed in Table I, we have two comparisons: injection A (large kick) with injection B (small kick), and injection C (large kick) with injection D (small kick). Most of the discussion will be focused on our findings for injections A (large kick) and B (small kick), and the purpose of performing injections C and D is to confirm the findings observed for injections A and B. In addition, we perform injection E with the goal of understanding the impact of the antisymmetric amplitude relative to the magnitude of the kick velocity. We include the relevant parameters of our injections in Table I.

We define the black-hole spin orientations at a reference frequency of $f_{\min} = 30$ Hz. Motivated by GW200129, the first binary with support for a nonzero kick velocity and spin precession, we fix the detector frame total mass value to match the most likely value inferred for this event: $M = 70 M_\odot$. We later repeat the same injections with $M = 45 M_\odot$ and $M = 25 M_\odot$. We choose three different SNR values: SNR = 26.8 (same as GW200129), 40, and 60. The luminosity distance is fixed such that the SNR has these values.

Most of the discussion is focused on the parameter inference of the face-on case, this is $\theta_{JN} = 0$. This is the angle between the total angular momentum and the line of sight of the observer. We later repeat the same injections with the angles $\theta_{JN} = \pi/4$ and $\theta_{JN} = \pi/2$.

Besides, in Table I we include the maximum value of the angle between the orbital angular momentum and the total angular momentum, commonly known as the opening angle $\beta$, which quantifies the amount of precession in the binary. When the opening angle is zero, there is no precession in the system. As we increase the opening angle, we allow the orbital plane to precess, and this causes mixing of various coprecessing GW harmonics to be present in the GW signal.

It is known that the presence of higher harmonics with different opening angles increases the measurability of precession [36]. To investigate the impact of the kick on parameter measurements and distinguish it from the impact of higher harmonics, we choose systems that suppress higher modes because of their orientation and mass symmetry. The latter suppresses all modes with odd $m$. Therefore, even when the orbital plane is precessing, the dominant $(\ell, m) = (2, 2)$ mode will not contain significant contributions from the corotating $(2, 1)$ mode. In contrast, for edge-on orientations, $\theta_{JN} = \pi/2$, several higher harmonics are expected to be more present in the signal. We later compare face-on to edge-on injections to understand the impact of kicks relative to that of higher harmonics.

We repeat each injection with IMRPhenomXPHM and IMRPhenomXO4a, to simulate the same binary with and without a kick. We use the same waveform model for injection and recovery to avoid introducing systematic errors arising from modeling differences between the injection and recovery models. This means the parameter biases that may arise in the Bayesian analysis can only be caused by waveform degeneracies and prior-induced constraints. These injections help us investigate the limitations of Bayesian parameter estimation when including or excluding certain physical effects in both the injected signal and the recovery model.

We use uniform priors in the masses distributed over $m_i = [1, 1000] M_\odot$. Spin magnitudes are uniformly distributed between $|\vec{\chi}_i| = [0, 0.99]$ and the spin orientations are isotropically distributed. The prior in the luminosity distance is uniform in [10, 10000] Mpc. All of our injections use a zero-noise realization while assuming a three-detector network formed by the two LIGO and the Virgo detectors at their design sensitivities [24,25].

To inject signals and estimate the marginalized posterior probabilities of the black-hole parameters, we employ Bilby [73], a Python-based GW-inference library, with the nested-sampling algorithm DYNESTY [74]. For the postprocessing of the posterior samples we use PESummary [75].

## III. INFERENCE OF SOURCE PARAMETERS IN THE PRESENCE OF A REMNANT KICK

### A. Effective precession parameter

We now compare the recoveries of the large-kick and small-kick injections. First, we look at the posterior distributions of the effective spin-precession parameter, $\chi_p$.

Figure 2 shows the posterior distributions of $\chi_p$ for injections A and B, which we refer to as "large kick" and "small kick" injections in the figure. These binaries are only different in their kick magnitude and their spin azimuthal angle $\phi_{12}$, and therefore have the same $\chi_p$ value.

When injecting and recovering the signals with IMRPhenomXO4a, we observe that the signal with a large kick has a more accurate recovery than the signal with a small kick. We find the same behavior with injections C and D. This suggests that the presence of a kick helps to extract more information from the signal. To test whether this is really true, we repeat the same injections simulating the signal with a model that does not include a kick or mode asymmetries.

When repeating the injections with IMRPhenomXPHM, without a kick and mode asymmetries in the signal, the Bayesian analysis infers that it is more likely that the source is an aligned spin system. For equal-mass face-on binaries, the signal looks very similar to that of a nonprecessing binary, and these results suggest that the signal does not include enough information for the analysis to determine precession. As IMRPhenomXPHM predicts a zero kick for both binaries, the only difference between injections A and B is $\phi_{12}$, which does not influence the $\chi_p$ value, and as expected, the $\chi_p$ recovery looks the same in both cases.

These results suggest that if the source experiences a kick, the GW signal is more informative and the Bayesian analysis is able to extract more information from the signal.

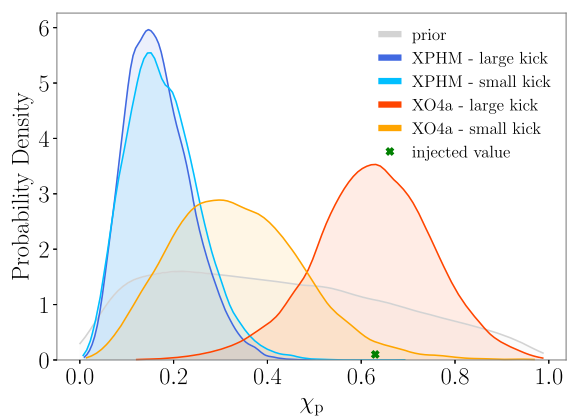

FIG. 2. Posterior probability distribution of $\chi_p$ for the injections A and B, which we refer to as the large kick and small kick injections, respectively. IMRPhenomXPHM posteriors are shown in blue, while IMRPhenomXO4a posteriors are shown in orange colors. The green cross indicates the injected value, and the prior distribution is shown in gray.

We observe that one can actually identify precession for equal-mass, face-on binaries through the improved measurement of the mass ratio and the spin tilt angles (see the following sections).

There are a few subtleties which make precession challenging to measure. Assuming a uniform prior in masses and spin magnitudes means that the prior in $\chi_p$ is not uniform (see Fig. 2). Therefore, all $\chi_p$ posteriors will exclude zero, and thus, every $\chi_p$ posterior could be considered as a measure of precession. In this line of thought, it is hard to know if the $\chi_p$ posterior represents a meaningful measurement of precession.

Since it is not always clear whether $\chi_p$ measurements represent meaningful precession measurements, a new parameter was proposed, which captures the observability of precession in a signal. This parameter is the precession SNR, $\rho_p$, and is based on the two-harmonic approximation proposed in [35,36]. It is defined as the SNR in the second most significant GW harmonic. A value of $\rho_p > 3$ represents a 1% false rate and is considered as strong evidence for the observability of precession.

We use the precession SNR to quantify the observability of precession in our injections. To compute this quantity we have used PESummary. For consistency with the parameter estimation analysis, we have used the same waveform models employed in parameter estimation and in the calculation of the precession SNR. We find that the posterior distributions are below the observability threshold in all face-on injections.

For the chosen equal-mass, face-on configurations, one can find that the two loudest harmonics are the $(\ell, m) = (2, \pm 2)$ coprecessing modes which get mixed into the $(2, 2)$ inertial mode (see Sec. III F for more details). PESummary does not consider the asymmetries between the $(2, 2)$ and $(2, -2)$ coprecessing modes included in IMRPhenomXO4a and identifies any subdominant harmonic except the $(2, \pm 2)$ as the second loudest harmonic. In most regions of the parameter space, such an assumption will not lead to significant bias. However, for large-kick face-on binaries, the mode asymmetries are non-negligible in the waveform and one should include them in the calculation of the precession SNR. As we incline the binary from face-on to edge-on, higher harmonics become louder and, thus, more relevant in the $\rho_p$ calculation. In fact, we find the PESummary calculations to be meaningful in the case of the edge-on injections.

Here, we want to use the same criterion to assess the parameter recoveries of all injections. Since neither $\chi_p$ nor $\rho_p$ might be meaningful parameters to quantify precession for the chosen binaries and waveform models, we prefer to look at the spin and mass ratio measurements directly.

### B. Spin tilt angle measurements

In the following, we look at the posteriors of the spin tilt angles whose priors are uncorrelated with other parameters. We mostly focus on the parameter recovery of IMRPhenomXO4a, as we can compare precessing binaries with significantly different kick magnitudes.

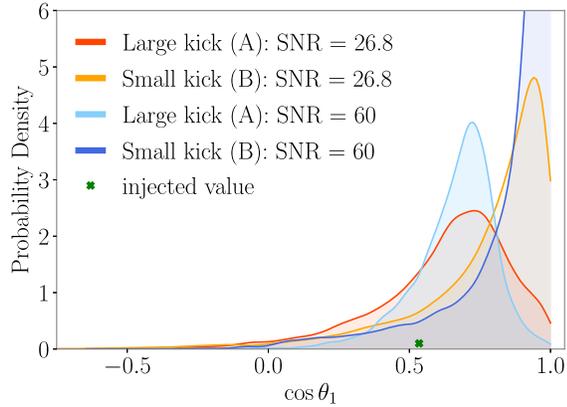

FIG. 3. Influence of the SNR on the posterior distributions of the tilt angles for the large-kick (A) and small-kick (B) injections with IMRPhenomXO4a. Orange colors represent posteriors with SNR = 26.8, while blue colors represent posteriors with SNR = 60. The injected value is shown with the green cross.

We observe that the tilt angles $\theta_1$ and $\theta_2$ are more accurately recovered in injections with a large kick than in the recovery of small-kick injections. All of the injections show the same trend. As an example, we include Fig. 3, which displays the posterior distributions of the primary tilt angle for two binaries that are only different in their kick magnitude. Instead of looking at $\theta_i$, we look at the cosine of the tilt angles $\cos\theta_i$, as the prior in $\theta_i$ is not uniform, while the prior in $\cos\theta_i$ is. Having a uniform (and uncorrelated) prior means that the posterior unambiguously represents the information extracted from the data.

Figure 3 shows that the primary spin tilt angle is more accurately recovered when there is a large kick. As we increase the SNR of the injected signal, the impact of the kick becomes significantly more visible. Figure 3 compares the recovery of the primary tilt for injections A (large kick) and B (small kick), for SNR = 26.8 (orange colors) and SNR = 60 (blue colors). We observe that the distinction between the large-kick and small-kick recoveries amplifies with the increase in SNR, which supports our hypothesis that the existence of a remnant kick leads to more precise spin measurements. In the following, we investigate whether the improvement in the accuracy of the measurements is truly correlated with the existence of a remnant kick in the signal.

#### 1. Importance of the observability of the merger

If the presence of a kick has any influence on the parameter recovery, then it is its imprint on the waveform that causes such differences in the measurements. As mentioned, the kick velocity is determined by the linear momentum that is radiated away as the two objects come closer together. Since most of the momentum is emitted in the last few orbits before the merger, the kick builds up during the merger, and leaves a non-negligible imprint on the merger phase of the waveform. To test the influence of the kick, we reduce its observable imprint by reducing the merger phase observable by the detector network.

We can reduce the observed merger phase by changing the total mass of the injected binary. This is because GW detectors are not equally sensitive in all frequency bins. The total mass of the binary determines the frequency range of the signal, and therefore which signal parts are observed by the detector. High-mass systems merge at low frequencies and have few cycles in band, while low-mass systems merge at high frequencies and have many more cycles in band. The most sensitive region of the LIGO and Virgo detectors is located between 100 and 400 Hz.

If the last few orbits of the merger occur in a frequency where the data is dominated by the detector noise, the SNR of the signal in the merger will be smaller than the SNR of the inspiral. Effectively, this could be thought of as calculating the linear momentum radiated away with a reduced final frequency that excludes the merger. Reducing the number of cycles of the merger observable by the detector network would reduce the linear momentum radiated away by the binary, and so the kick magnitude. We can call the kick velocity observable by the detectors as the *effective kick velocity*, $v_{\text{eff}}$.

By diminishing the kick imprint on the injected signal we expect to decrease the effect of the kick on the spin recovery. If our hypothesis is true, we expect the decrease in total mass to increase the bias on the recovery of the spin posteriors. In binaries with no significant kick, decreasing the total mass will not change the kick imprint, as it is already small, so we expect the spin posteriors to remain unchanged. The injections we have performed, which are included in Table I, have a total mass of $M = 70 M_\odot$. With such value, the merger phase occurs between 100 to 300 Hz, which is exactly the frequency range where the detectors are the most sensitive. We repeat the same injections with total mass values of $M = 25 M_\odot$ and $M = 45 M_\odot$, for which the merger occurs at higher frequencies where the detectors are not as sensitive. For the injections with $M = 25 M_\odot$, the merger occurs at around $10^3$ Hz, while for the injections with $M = 45 M_\odot$, at around 500 Hz.

We find that in the presence of a large kick, decreasing the total mass leads to increased bias, as the posteriors shift away from the injected value. In Fig. 4 we show the posterior distributions of the spin tilt for the large-kick (A) and small-kick (B) injections with different total mass values. With the decrease in total mass, the posteriors shift towards aligned-spin configurations and align with the posterior of the small-kick injection (B). When the binary has a small kick, we observe that changes in the total mass do not impact significantly the spin posteriors. We find the same behavior for both primary and secondary spin tilt angles.

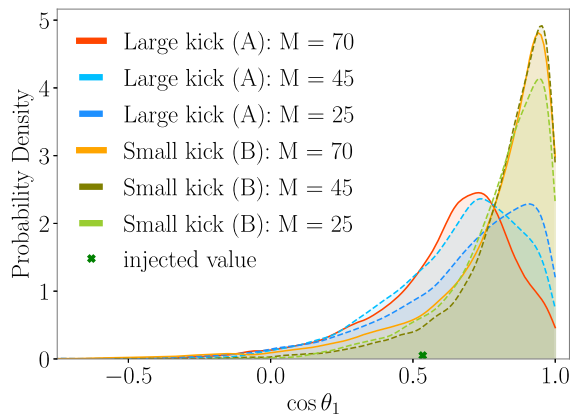
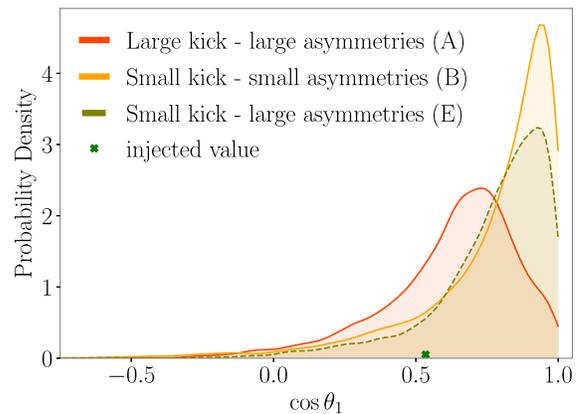

FIG. 4. Influence of the total mass on the posterior distributions of the tilt angles for the large-kick (A) and small-kick (B) injections with IMRPhenomXO4a. We test three different total mass values: M = 25, 45 and $70 M_\odot$. The injected value is indicated with the green cross.

FIG. 5. This figure compares the posteriors of the primary spin tilt of injections A (a binary with a large kick and large asymmetries), B (a binary with a small kick and small asymmetries), and E (a binary with a small kick and large asymmetries), injected and analyzed with IMRPhenomXO4a. This plot shows the impact of the kick magnitude relative to the impact the antisymmetric amplitude on the spin measurements.

Our results confirm that observing the merger phase is key to making more accurate measurements, only in the presence of a significant kick. We know the kick is described by the symmetric and antisymmetric parts of the waveform, and in the next two subsections we investigate how each of these play a role in the parameter inference.

### 2. Impact of the mode asymmetries

The authors of Ref. [41] already quantified the impact of including mode asymmetries when inferring in-plane spins for binaries with $q = 2$ and SNR = 100. By looking at binaries with the same antisymmetric amplitude, but different kick magnitudes, they found that it is only the asymmetry content what causes the improved recoveries. Here, we rediscuss this matter for the case of equal-mass binaries with SNR = 26.8.

Based on the relation between kick magnitudes and the antisymmetric waveform (see Fig. 1), we find that there are three possible cases: (i) large kicks with large asymmetries, (ii) small kicks with small asymmetries, and (iii) small kicks with large asymmetries. Injections A, B, and E represent these three cases, respectively. Regarding the spin parameters, these injections have the same spin magnitude and tilt angles and are only different in their $\phi_{12}$ value (see Table I).

Figure 5 compares the spin posteriors of these three injections. When we compare the injections with the same kick magnitude but different antisymmetric amplitude, B against E, we observe that the magnitude of the asymmetries has a small impact on the spin recovery. In the same way, when we compare the injections with the same antisymmetric amplitude but different kick magnitude,

A against E, a large kick magnitude leads to more accurate posteriors. In summary, we find that it is the combination of a large antisymmetric amplitude with a large kick magnitude in the GW signal which helps make more accurate measurements.

In addition, we investigate the impact of including mode asymmetries when analyzing a signal with a large kick. To do so, we repeat the analysis using the full IMRPhenomXO4a (including asymmetries) for the injection, and recovering the signal with a version of IMRPhenomXO4a that excludes the asymmetries. As displayed in Fig. 6, we observe that the posteriors obtained excluding mode asymmetries in the model are slightly less precise, but generally very similar to the posteriors obtained when including them. We observe the same behavior for the secondary spin tilt.

These results suggest that it is mostly the additional structure in the observed signal that is responsible for an improved spin measurement, even if the model used for the recovery does not include the mode asymmetries. The additional phase information from the asymmetries in the injected signal helps the analysis to disfavor small spin tilts. As we increase the SNR to 60, the analysis that excludes mode asymmetries has significantly more biased posteriors. This observation is in agreement with Ref. [41], where the SNR was fixed to 100 such that the parameter biases were visible.

We now deactivate the mode asymmetries in the injected signal and test whether the asymmetries are fully responsible for the improved recoveries or not. We repeat these injections with IMRPhenomXO4a, excluding the mode asymmetries both in the injection and recovery models. We compare the posteriors with those of IMRPhenomXPHM, since IMRPhenomXPHM also excludes mode asymmetries, but has a slightly less accurate description of the merger and ringdown phases.

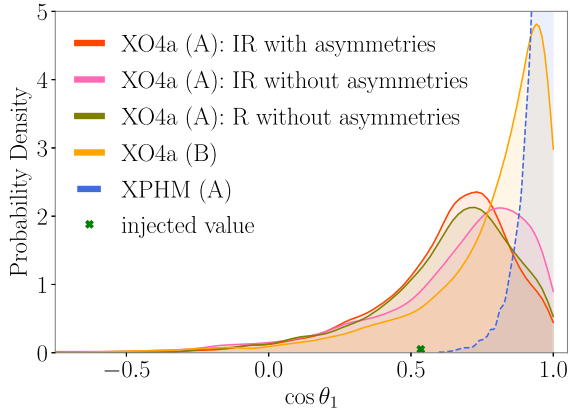

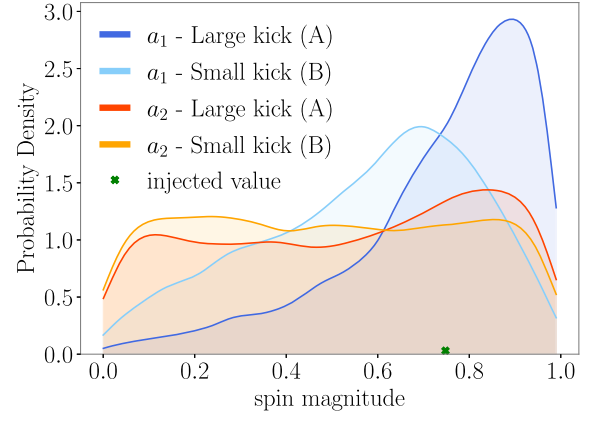

FIG. 6. This figure shows the posterior distributions of the primary spin tilt angle of injection A analyzed with IMRPhenom-XO4a in red and with a version of IMRPhenomXO4a that excludes mode asymmetries in the injection and recovery models in pink. The acronyms of the legend "I" and "R" stand for injection and recovery, respectively. In brown, we show the posterior of injection A, where the injected signal includes asymmetries, and the recovery model excludes them. In blue, we show the posterior of injection A injecting and recovering the signal with IMRPhenomXPHM. The figure also includes the posterior of the primary spin tilt angle of the small-kick injection B (orange) analyzed with IMRPhenomXO4a.

Figure 6 shows that the posterior of IMRPhenomXO4a without mode asymmetries (pink) is less accurate than the analysis with the full IMRPhenomXO4a model (red). However, the IMRPhenomXO4a posterior obtained without asymmetries is still more accurate than the low-kick injection analyzed with the full IMRPhenomXO4a. It is also significantly more accurate than the IMRPhenomXPHM posterior. This suggests that the modeling of the (symmetric) merger-ringdown phase plays a role in describing the imprint of the kick and impacts the parameter recoveries. We find that it is the combination of the inclusion of the mode asymmetries and the more accurate modeling of the merger-ringdown phases which ultimately lead to the observed spin measurements of IMRPhenomXO4a.

### C. Spin magnitude measurements

We now look at the measurements of the spin magnitudes, $a_1$ and $a_2$. In the same way as for the spin tilt angles, we observe that the recovery in injections with large kicks is in general more accurate than in injections with small kicks.

Figure 7 includes the posterior distributions of the two spin magnitudes and compares the posteriors obtained for injections A (large kick) with B (small kick). In general, we find that $a_2$ is poorly constrained for both large-kick and small-kick injections, with the large-kick recovery being slightly more accurate. In the case of $a_1$, however, we observe significantly different posterior distributions for the large-kick and small-kick injections.

FIG. 7. Posterior distributions of the spin magnitudes $a_1$ (blue colors) and $a_2$ (orange colors) for the large-kick (A) and small-kick (B) injections with IMRPhenomXO4a.

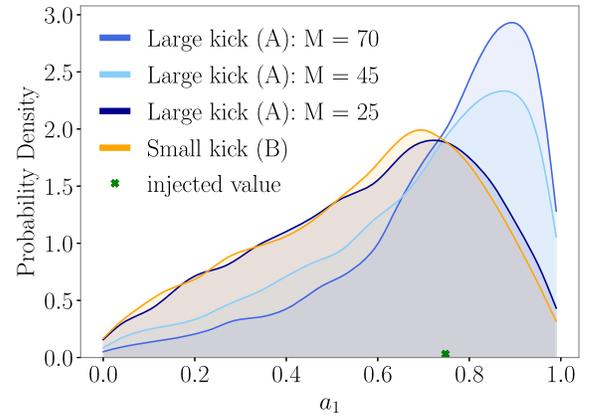

FIG. 8. Posterior distributions of the primary spin magnitude for the large-kick injection A (blue) and the small-kick injection B (orange) with IMRPhenomXO4a. The plot shows the influence of reducing the total mass on the primary spin magnitude.

We investigate whether the fact that the recovery of the primary spin magnitude is more accurate in the large-kick cases is correlated with the kick magnitude. In the same way as for the spin tilts, we reduce the total mass of the binary to reduce the kick imprint observable by the detectors. When reducing the total mass, the recovery of the primary spin magnitude spreads out and becomes less precise than the original posterior with $M = 70 M_\odot$. See Fig. 8 for the case of injection A. This supports our idea of the imprint of the kick leading to more accurate measurements of the spins.

### D. Spin azimuthal angle measurements

Figure 9 shows the posterior distributions of the spin azimuthal angle $\phi_{12}$ of injections A and B. For SNR = 26.8, both posteriors appear flat and uninformative.

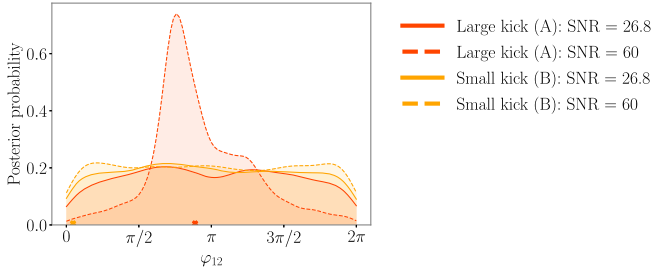

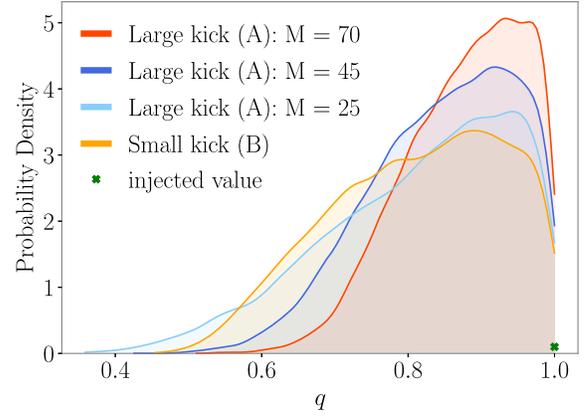

FIG. 9. Posterior distributions of the spin azimuthal angle $\phi_{12}$ of the large-kick injection A (red) and the small-kick injection B (orange) with IMRPhenomXO4a. The injected values are different in each injection and are indicated in the plot with colored crosses. Solid lines indicate the posteriors with SNR = 26.8, while the dashed lines indicate the posteriors with SNR = 60.

However, as we increase the SNR to 60, we observe that the posterior of the large-kick injection is peaked and centered around the injected value at $\phi_{12} \sim \pi$. On the other hand, the small-kick injection remains uninformative even when increasing the SNR to 60. This suggests once again that a source that experiences a remnant kick radiates a signal that is more informative than if the source experienced no remnant kick.

### E. Mass ratio measurements

We have seen that having a significant kick leads to more accurate measurements of the effective spin-precession parameter $\chi_p$, the spin magnitude, tilt, and azimuthal angles. Apart from the individual spins, the effective spin-precession parameter $\chi_p$ also depends on the mass ratio [see Eq. (6)]. Here we investigate whether the kick has an impact on the mass ratio measurement.

In the same way as for the spins, we find that large-kick injections have more accurate recoveries of the mass ratio than small-kick ones. Figure 10 shows the posterior distributions of the mass ratio for injections A (large kick) and B (small kick) with SNR = 26.8.

As shown in Fig. 10, when decreasing the observed strength of the merger by decreasing the injected total mass, the uncertainty in the mass ratio posteriors increases for the large kick injections. This points to the importance of observing the merger, which contains the imprint of the kick.

### F. Discussion

Our results support that the imprint of the kick leads to more precise measurements of the intrinsic properties. We believe that this observation is connected to the measurability of the mode asymmetries. In our case, since we are using the model IMRPhenomXO4a, these are restricted to the dominant mode asymmetry.

When equal-mass binaries are oriented face-on to the detectors, all subdominant inertial harmonics are suppressed in the signal, which means that only the $(\ell, m) = (2, 2)$ inertial harmonic can be measured. The dominant inertial harmonic can be expressed in terms of the coprecessing harmonics based on a frame rotation [63],

$$h_{2,2} = \sum_{m' \in \pm\{1,2\}} h^{CP}_{2,m'} e^{i2\alpha} d^2_{m',2}(-\beta) e^{-im'\epsilon}, \quad (13)$$

where $d^\ell_{m',m}$ is the Wigner matrix that depends on the opening angle $\beta$ between the total and orbital angular momentum, and $\alpha$ and $\epsilon$ are the remaining two angles that are required to describe the instantaneous orientation of the orbital plane. The leading-order amplitude of the $h^{CP}_{2,1}$ harmonic is proportional to the mass difference between the two objects. So for equal-mass binaries, this harmonic is suppressed. The dominant inertial harmonic is thus a combination of the $(\ell, m) = (2, \pm 2)$ coprecessing harmonics:

$$h_{2,2} = h^{CP}_{2,2} e^{i2\alpha} d^2_{2,2}(-\beta) e^{-i2\epsilon} + h^{CP}_{2,-2} e^{i2\alpha} d^2_{-2,2}(-\beta) e^{i2\epsilon}. \quad (14)$$

FIG. 10. Posterior distributions of the mass ratio for the large-kick (A) and small-kick (B) injections with IMRPhenomXO4a. The figure also shows the influence of the total mass on the mass ratio posterior of injection A, as we test three different total mass values: $M = 25$, 45 and $70M_\odot$. The injected value is indicated with the green cross.

According to the two-harmonic approximation, two harmonics with different $\beta$ dependencies need to be observed to unambiguously measure precession [35]. These can be two harmonics in the inertial frame, or two harmonics in the coprecessing frame that mix into one inertial harmonic, as it was later shown in [42]. When using a waveform model that assumes the symmetry relation $h^{CP}_{2,2} = h^{*CP}_{2,-2}$, then the inertial harmonic $h_{2,2}$ is the combination of the coprecessing $h^{CP}_{2,2}$ harmonic with its complex conjugate. This means we cannot measure two harmonics with different $\beta$ dependencies and, therefore, precession is not measurable. This is the case

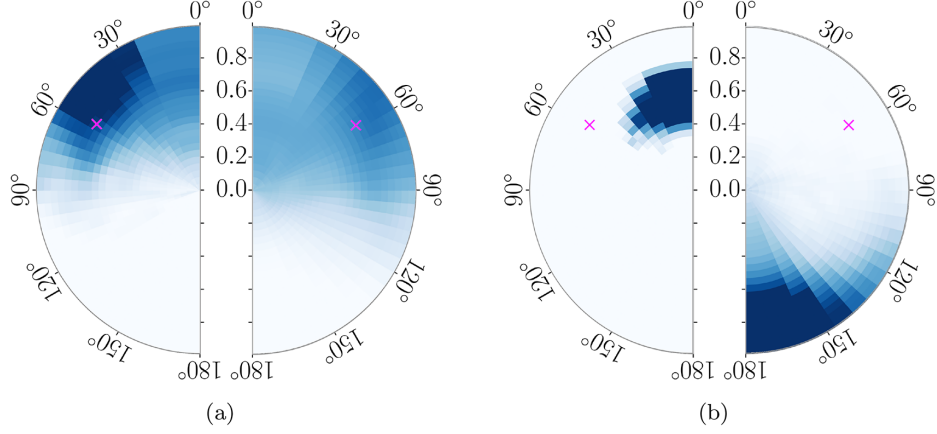

FIG. 11. Two-dimensional spin posterior distributions of the large-kick injection A with IMRPhenomXO4a (a) and IMRPhenomXPHM (b) for SNR = 26.8. In each case, the panel on the left shows the posterior distribution of the primary spin, while the panel on the right shows the posterior of the secondary spin. Each panel displays the posterior of the spin magnitude and the spin tilt. Injected values are indicated with pink crosses in the plots.

for all SEOBNR and Phenom waveform models, except IMRPhenomXO4a.

However, if the waveform model includes the asymmetry between the $h^{CP}_{2,2}$ and $h^{CP}_{2,-2}$ harmonics, then the inertial harmonic $h_{2,2}$ is indeed a combination of two coprecessing harmonics with different $\beta$ dependencies. Therefore, precession will be observable if the two dominant coprecessing harmonics are sufficiently strong in the signal, or equivalently, the antisymmetric part of the waveform is observable. A GW signal with a significant kick necessarily includes large mode asymmetries and, therefore, based on the previous argument, we can expect to measure precession more accurately than when the mode asymmetries are small or actually neglected.

Since IMRPhenomXPHM assumes the symmetry relation between the negative- and positive-$m$ coprecessing harmonics, effectively, the radiated signal only contains one harmonic, and the signal appears to be similar to that of a nonprecessing binary. Therefore, we do not expect precession to be measurable with IMRPhenomXPHM for face-on, equal-mass binaries.

As expected from the previous argument, the parameter recovery of IMRPhenomXPHM appears to be biased in several parameters: $\chi_p$ as shown in Fig. 2, and the spin tilt angles as shown in Fig. 11. IMRPhenomXPHM shows support for the primary spin being aligned with the orbital angular momentum and the secondary being antialigned with the orbital angular momentum. The results are consistent with the expectation of the two-harmonic approximation: face-on, equal-mass precessing waveforms appear as nonprecessing waveforms. In the case of IMRPhenomXPHM, the sampler chooses points in the parameter space that are consistent with aligned-spin configurations.

The bias in the spin orientations appears to be compensated with unequal mass ratios, as shown in Fig. 12. While the posterior distributions of IMRPhenomXO4a shift toward the injected value as we increase the signal SNR to 60, we find that the IMRPhenomXPHM posteriors do not become more accurate. The mass ratio estimates remain biased.

Such bias is not unexpected. The face-on precessing signal mimics the signal of a nonprecessing binary. However, the phase evolution is not the same as the one from a fiducial binary where the in-plane spin components have simply been set to zero. Precession adds a secular phase drift of twice the precession phase (i.e., the accumulated phase of the orbital angular momentum around the total angular momentum) [35,76]. This missing phase can be compensated by modifying the intrinsic parameters, most notably the mass ratio in the IMRPhenomXPHM analysis. The IMRPhenomXO4a analysis does not suffer from this bias as the signal contains more structure due to mode asymmetries and an updated ringdown description. Those additional effects can only be mimicked by actually converging towards the correct properties of a precessing binary.

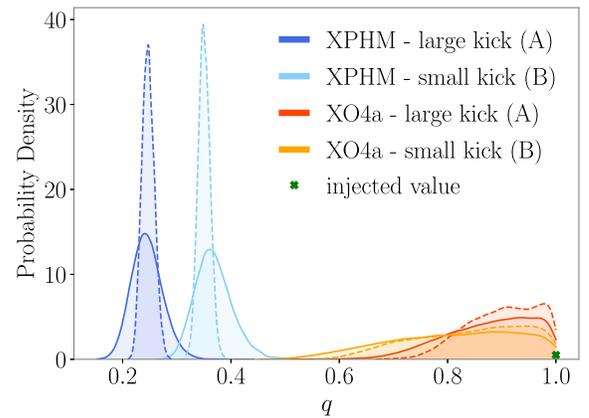

FIG. 12. Posterior distributions of the mass ratio for injections A (large kick) and B (small kick) with IMRPhenomXPHM (blue colors) and IMRPhenomXO4a (orange colors). Solid lines indicate the posteriors of the injections with SNR = 26.8, and dashed lines indicate the posteriors of the injections with SNR = 60.

We further observe that the measurement of the mass ratio influences the $\chi_p$ posterior of IMRPhenomXPHM. By using an analytical expression of the $\chi_p$ prior (see Sec. IV B in [42]), we can find how the prior distribution changes when fixing the mass ratio. We use the median value of the mass ratio posterior of A, $q = 0.24$, and we find that the new $\chi_p$ prior distribution has a peak around $\chi_p \approx 0.15$, which is exactly where the median of the A IMRPhenomXPHM distribution lies.

As we inject and recover with the same signal model, and use a zero noise realization, one might still expect the analysis to converge on the injected parameters. After all, the point of highest likelihood is, by construction, where the injected parameters are. However, the Bayesian analysis converges on the highest posterior density, which is determined by both the prior volume and the likelihood. If the prior density is small for our injection (i.e., we have chosen a special point in the parameter space) and the likelihood remains high across a significant parameter-space volume for nonprecessing binaries, the interpretation as a nonprecessing binary will be favored unless the SNR is extremely high. Here we do not explore very high SNRs, so we remain in the biased regime. For further comparisons between the two analyses, we include corner plots with posteriors of the most important injections in the Appendix.

### G. Impact of the inclination angle

As we incline a binary from face-on to edge-on orientations, we expect to observe multiple GW harmonics, which generally help determine the source properties more accurately. When discussing the observability of precession in the case of large-kick injections, we also need to consider the impact of the inclination angle on the observability of the mode asymmetries, namely, the dominant-mode asymmetries as included in IMRPhenomXO4a. In the same way as the (2, 2) mode, the amplitude of the antisymmetric waveform reaches its maximum face-on and it decreases as we incline the system to edge-on [53].

When injecting and recovering signals with IMRPhenomXPHM, the posteriors shift towards the injected value as we incline the system, and as expected from [42], the model recovers the parameter best for edge-on inclinations (see Fig. 13). In the case of IMRPhenomXO4a, however, we observe that when the binary has a large kick and in turn large asymmetries, the $\chi_p$ recovery becomes less accurate as we incline the system. For an inclination of $\theta_{JN} = \pi/4$, the posterior looks similar to the face-on case. For $\theta_{JN} = \pi/2$, we observe that the $\chi_p$ recovery becomes less accurate than for $\theta_{JN} = 0$ and $\theta_{JN} = \pi/4$ inclinations and is similar to the IMRPhenomXPHM recovery. As the amplitude of the asymmetries decreases with the inclination and it vanishes for edge-on inclinations, these results show that it is the observability of the asymmetries which helps determine precession in the case of the $\theta_{JN} = 0$ and $\theta_{JN} = \pi/4$ injections. In the case of small kick injections and small mode asymmetries, IMRPhenomXO4a recovers precession more accurately when increasing the inclination angle, in the same way as with IMRPhenomXPHM.

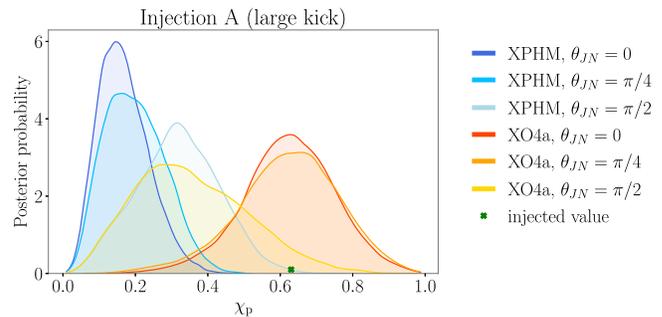

FIG. 13. Posterior probability distribution of $\chi_p$ for the injection A (large kick) with an inclination angle of $\theta_{JN} = 0, \pi/4$ and $\pi/2$ (rad). IMRPhenomXPHM posteriors are shown in blue, while IMRPhenomXO4a posteriors are shown in orange colors. The green cross indicated the injected value.

### H. The case of GW200129

The source of GW200129 is thought to be a binary black hole with a mass ratio of $q = 0.6^{+0.4}_{-0.2}$, an inclination of $\theta_{JN} = 0.5^{+0.3}_{-0.3}$ [30], and a remnant kick velocity of $v \sim 1542^{+747}_{-1098}$ km/s [29]. In both studies the signal was analyzed with the waveform model NRSur7dq4, which incorporates mode asymmetries. We have shown in Sec. II D that the presence of a large kick necessarily implies the existence of mode asymmetries in the signal, independent of the mass ratio. Indeed, Ref. [41] showed how considering mode asymmetries in the inference of the source properties was essential in finding precession in the system. In addition, we have shown that the GW imprint of the kick can help to extract more meaningful information about the spins in equal-mass binaries. With the inferred mass ratio and inclination values of GW200129, the signal probably contains higher harmonics, which help identify precession. Based on our study we find it is plausible that the presence of a kick, and therefore mode asymmetries, helped determine precession in the signal.

### IV. KICK MEASUREMENTS

We infer the kick posterior of each injection using the posteriors of the source properties, and we investigate whether one can distinguish a signal with a large kick from a signal with a low kick. We should note that our findings on the imprint of the kick leading to more accurate measurements are independent of our ability to constrain the kick velocity from the signal. We might not be able to constrain a kick as for GW200129, but the presence of a kick would still have an impact on parameter estimation.

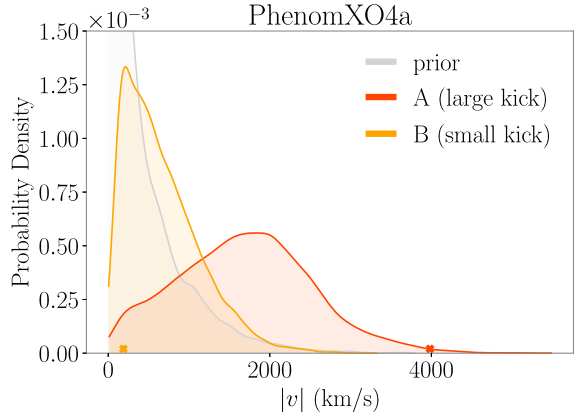

FIG. 14. Posterior distributions of the kick magnitude for the injections A (large kick) and B (small kick) computed with IMRPhenomXO4a. The prior distribution is included in gray color, and the crosses in the $x$ axis indicate the kick magnitude of each of the injected signals.

However, if more precise spin measurements are possible in the presence of a large kick, then the decrease in the uncertainty of the spin measurements should lead to more informative kick posteriors.

Figure 14 includes the kick posteriors of the face-on injections we have performed with IMRPhenomXO4a. We observe that the posterior distribution of the large-kick injection A (with $v \sim 4000$ km/s) is informative and has a distinctive shape different from the prior distribution. Even though its median value is not centered at the injected value, its value is only consistent with a precessing binary. In the case of low-kick injection, B, the kick prior has a dominant effect on the posterior distributions. As expected, the ability to recover the injected value improves as we increase the signal SNR. As shown in Fig. 15, the increase in SNR leads to more accurate kick posteriors with a reduced support for zero kicks. On the other hand, IMRPhenomXPHM cannot predict large precessing kicks and recovers, in all cases, a low kick velocity close to the injected value.

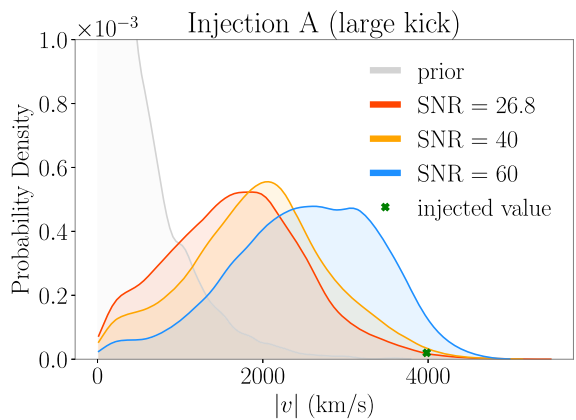

FIG. 15. Influence of the signal SNR on the kick posteriors of injection A estimated with IMRPhenomXO4a. The green cross indicates the injected kick value. The plot shows that the increase in SNR helps in reducing the support for zero kicks on the inferred kick posterior and leads to more accurate constraints of the true value.

We further investigate what leads to an accurate kick posterior, whether it is the parameter estimation samples and/or the mapping from intrinsic properties to the kick velocity. First, we test the influence of the posterior samples. We use samples estimated with and without including mode asymmetries in IMRPhenomXO4a, and map them to kick velocities using the complete IMRPhenomXO4a model, which includes mode asymmetries. Figure 16 shows the comparison of these two kick posteriors for the large-kick injection A. We can see that including mode asymmetries on the inference of the source parameters plays an important role and impacts the kick posterior.

Second, we test the impact of including mode asymmetries on the model used to infer the kick velocity. Since mode asymmetries are responsible for the out-of-plane component of the kick velocity, we find that without asymmetries one can only obtain kick magnitudes up to 500 km/s. Hence, in cases where the injected kick value is a large precessing kick, the model will not be able to recover it. This states the importance of including mode asymmetries in the estimation of the kick posterior.

Besides, as mentioned before, the same binary configuration might have two different estimates depending on the waveform model that is used. This means that using two different models, one for the parameter estimation and a different one for the kick inference, can introduce large systematic errors, as the kick estimate of the posterior sample obtained with model 1 might not agree with the kick estimate of model 2. However, it is common practice to use posterior samples obtained with PHENOM and SEOBNR waveform models and map them to remnant kicks with either NRSurrogate fits or NR fitting formulas (see, e.g., [18,77,78]). We would like to emphasize the importance of using the same waveform model both for parameter estimation and the kick inference, to avoid introducing systematic errors and make more meaningful statements about the kick velocity.

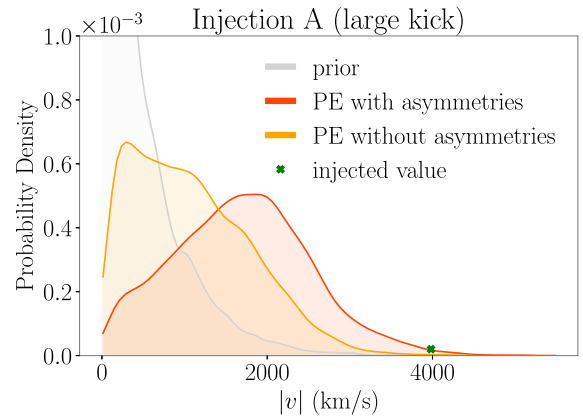

FIG. 16. Kick posterior distributions of injection A displaying the impact of excluding mode asymmetries on the recovery model of the parameter estimation analysis. To obtain the kick posterior in orange we used posterior samples obtained excluding mode asymmetries in IMRPhenomXO4a, and estimated the kick posterior using the complete model, which includes mode asymmetries. The posterior in red represents the kick posterior obtained using the complete IMRPhenomXO4a model for both parameter estimation and the kick estimate. The green cross indicates the injected kick value.

## V. CONCLUSIONS

We have investigated whether the GW imprint of the kick can help to extract more information from the signal. To exclude the impact of higher harmonics in parameter estimation, we have focused on face-on, equal-mass binaries. This is also motivated by the fact that the largest kick velocities are found in equal-mass binaries. Measuring spin precession in these orientations and mass ratios has been shown to be challenging. Here, we have explored whether the presence of a kick in the GW signal can improve our ability to infer precession in these cases. Our findings are summarized in the following bullet points:

(i) In Sec. II D, we have presented the relation between the dominant mode asymmetry and the kick magnitude. We explore the two-spin parameter space and find that large mode asymmetries do not necessarily induce large kicks. However, for a given mass ratio and spin misalignment, large kicks are only generated by signals with large asymmetries.

(ii) Based on our injection and recovery study, we observe that we can make more accurate measurements of the spins and mass ratio when the GW signal includes a large kick. This applies to equal-mass, face-on binaries, for which the merger phase of the signal is clearly visible by the detectors. Since the kick leaves an imprint on the merger phase of the GW signal, the impact of the kick is the largest when the merger phase is observable by the detectors.

(iii) We find that one can distinguish between precessing and nonprecessing binaries in signals from equal-mass, face-on binaries, only when the signal includes a kick. In Sec. III F, we have discussed the problem in the context of the two-harmonic approximation. This is a framework that quantifies the measurability of precession and it states that to unambiguously determine precession, two different GW modes need to be observable. Signals from equal-mass, face-on binaries only include the dominant mode, which makes precession difficult to be measured. However, if these signals include large kicks, they necessarily include large asymmetries. The dominant mode can be decomposed into the two coprecessing dominant modes, meaning that, if mode asymmetries are included in the signal, as in IMRPhenomXO4a, the dominant mode asymmetry can make the signal more informative. The presence of mode asymmetries in the signal can help to observe two harmonics and to identify precession. If we exclude mode asymmetries in the description of the large-kick signal, as with IMRPhenomXPHM, then the two coprecessing dominant modes will be symmetric, and the signal will only contain one mode, making precession hard to measure. In the case of a small kick where the asymmetries are small, the dominant coprecessing modes are close to being symmetric, for which identifying precession will again be challenging.

(iv) We find that the formulation of the two-harmonic approximation is not complete for the case of face-on, equal-mass binaries, as it does not consider the asymmetries between the $(\ell, m) = (2, 2)$ and $(\ell, m) = (2, -2)$ GW harmonics. As many observed GW candidates are expected to be close to this specific configuration, we find it is important to consider the asymmetries in the dominant mode to estimate the observability of precession in current GW events.

(v) Regarding kick measurements, we find that the waveform model used in the Bayesian analysis is equally important as the model used for the estimation of the kick posterior. We observe that mixing different waveform models can introduce systematic errors into the kick posterior estimate. In addition, we find that including mode asymmetries is essential to infer an accurate kick posterior.

## ACKNOWLEDGMENTS

We are grateful to Mark Hannam, N. V. Krishnendu, Maite Mateu-Lucena, and Vijay Varma for useful discussions. We also thank the anonymous referee for the valuable comments on this paper. This work was supported by the Max Planck Society's Independent Research Group Grant. Computations were carried out on the Holodeck cluster of the Max Planck Independent Research Group "Binary Merger Observations and Numerical Relativity."

## APPENDIX

For completeness, we include corner plots with the posteriors of injections A (large kick) and B (small kick) using IMRPhenomXO4a and IMRPhenomXPHM. In Figs. 17–20, we show the posteriors of the injections with SNR = 26.8, while Figs. 21 and 22 show posteriors of the IMRPhenomXO4a injections with SNR = 60. The corner plots include posteriors of the chirp mass, the mass ratio, the component spin magnitudes, the spin tilts, the spin azimuthal angle $\varphi_{12}$, the effective spin $\chi_{\text{eff}}$, the effective precession parameter $\chi_{\text{p}}$, and the luminosity distance. The vertical lines in dark blue indicate the true parameter values.

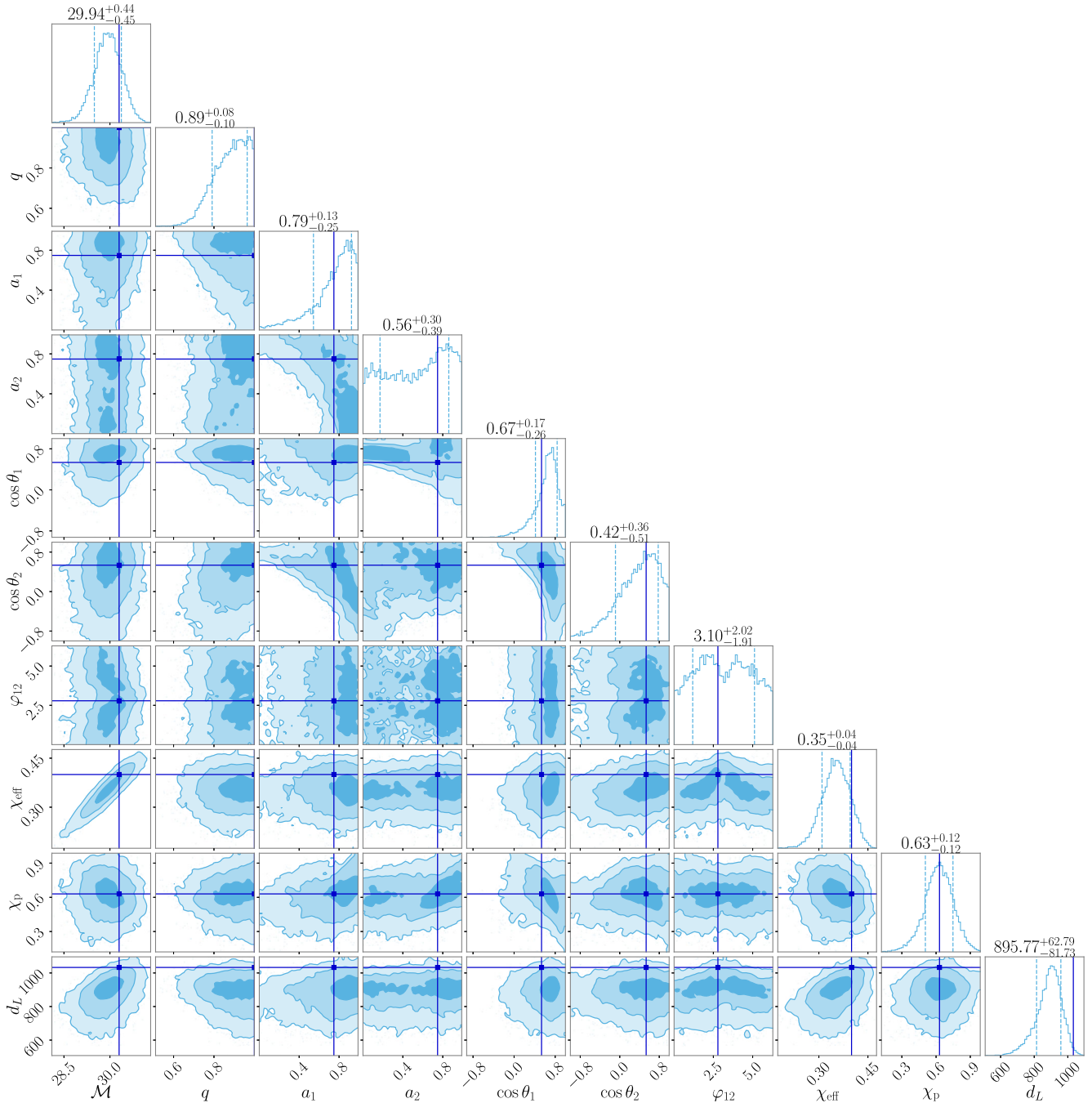

FIG. 17. Corner plot of the posteriors of injection A using IMRPhenomXO4a with SNR = 26.8. The plot includes the posteriors of the chirp mass, the mass ratio, the component spin magnitudes, the spin tilts, the spin azimuthal angle $\varphi_{12}$, $\chi_{\text{eff}}$, $\chi_{\text{p}}$, and the luminosity distance. The true parameter values are indicated by the vertical lines in dark blue.

When comparing the two IMRPhenomXO4a injections in Figs. 17 and 18, we observe that the posteriors of the chirp mass $\mathcal{M}$ and the luminosity distance $d_L$ are slightly biased in the large kick injection (A). This is because signals with significant kicks have amplitude modulations in the merger phase caused by the anisotropic emission of GWs. In the case of a binary with the parameters of injection A, the remnant moves towards the observer nearly in the line of sight. This means that the signal has a slightly smaller amplitude than if the remnant moved away from the observer, or if the binary had no remnant kick. A smaller

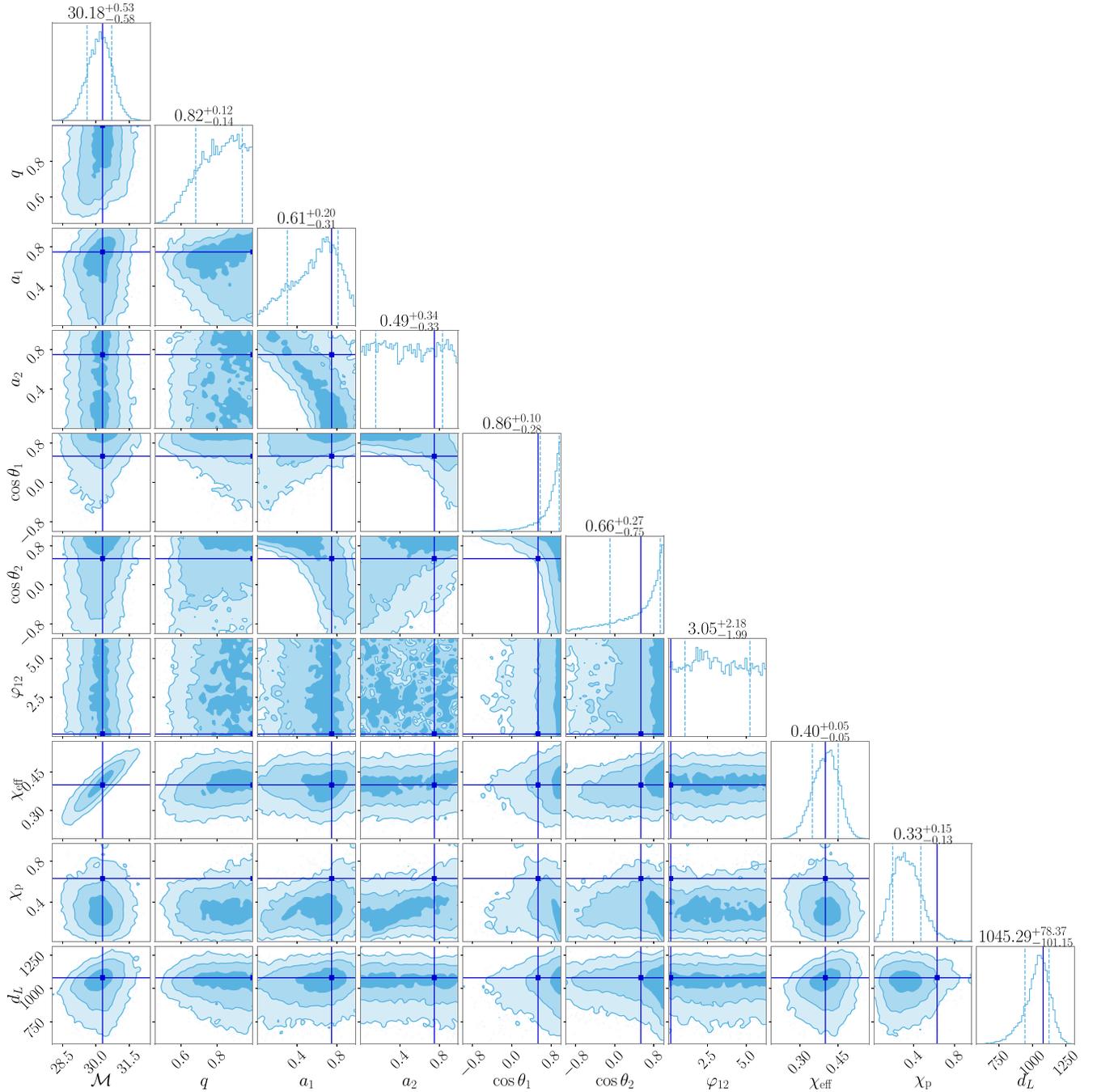

FIG. 18. Corner plot of the posteriors of injection B using IMRPhenomXO4a with SNR = 26.8. The plot includes the posteriors of the chirp mass, the mass ratio, the component spin magnitudes, the spin tilts, the spin azimuthal angle $\varphi_{12}$, $\chi_{\rm eff}$, $\chi_{\rm p}$, and the luminosity distance. The true parameter values are indicated by the vertical lines in dark blue.

GW amplitude can be mimicked by a less massive binary, hence the bias in chirp mass. To compensate the amplitude in the inspiral corresponding to a lower-mass binary, the luminosity distance needs to be modified such that it is closer to the observer. The bias reduces as we increase the SNR (see Fig. 21). On the other hand, we observe that the spin posteriors of the small-kick injection (B) remain biased even when increasing the SNR to 60, while those of the large-kick injection become more accurate with the increase in SNR.

As discussed in Sec. III F, we generally observe larger biases in IMRPhenomXPHM than in IMRPhenomXO4a posteriors. These differences are visible in Figs. 17–20 for a number of source parameters, including $\chi_{\rm p}$.

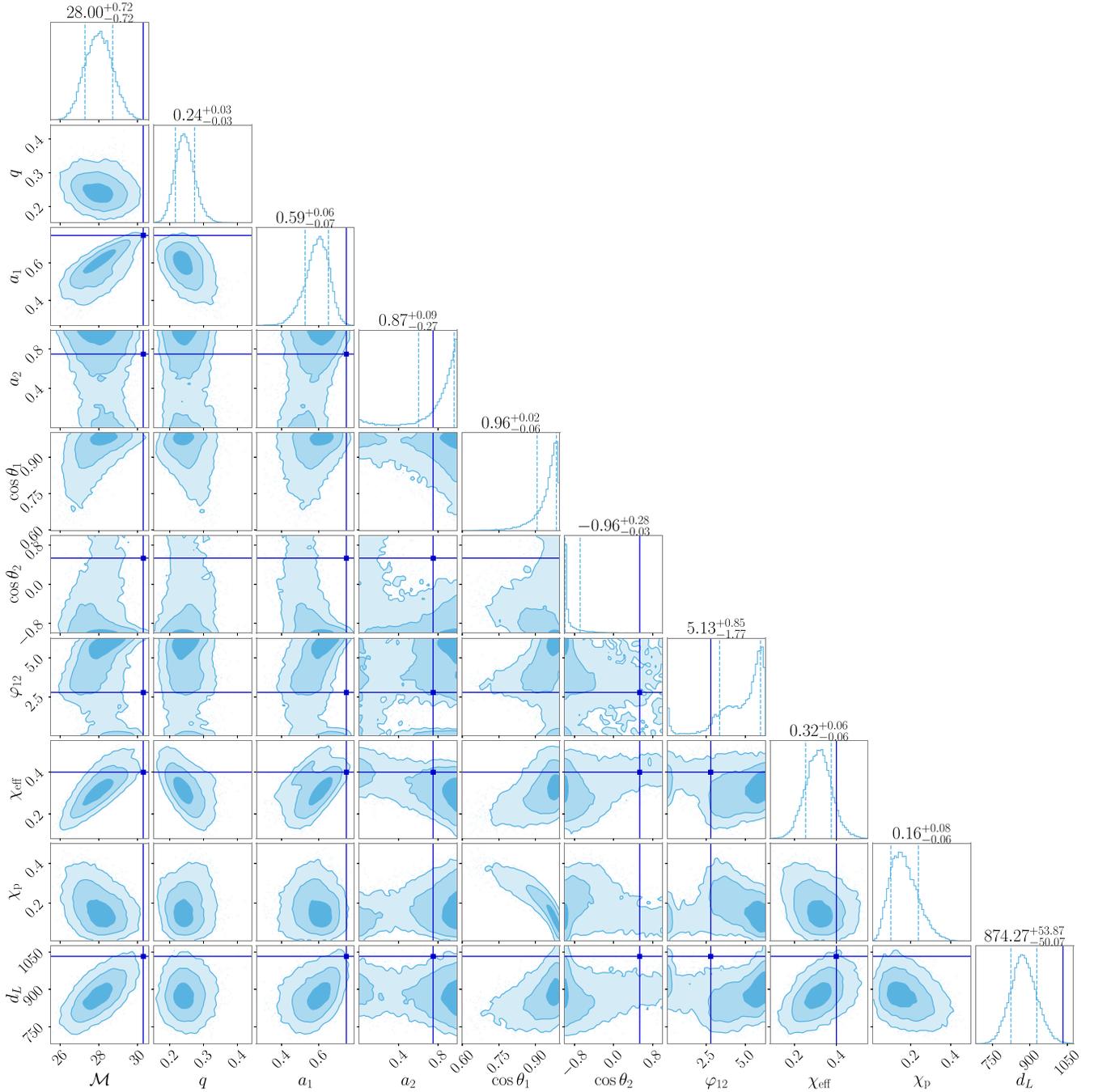

FIG. 19. Corner plot of the posteriors of injection A using IMRPhenomXPHM with SNR = 26.8. The plot includes the posteriors of the chirp mass, the mass ratio, the component spin magnitudes, the spin tilts, the spin azimuthal angle $\varphi_{12}$, $\chi_{\text{eff}}$, $\chi_{\text{p}}$, and the luminosity distance. The true parameter values are indicated by the vertical lines in dark blue.

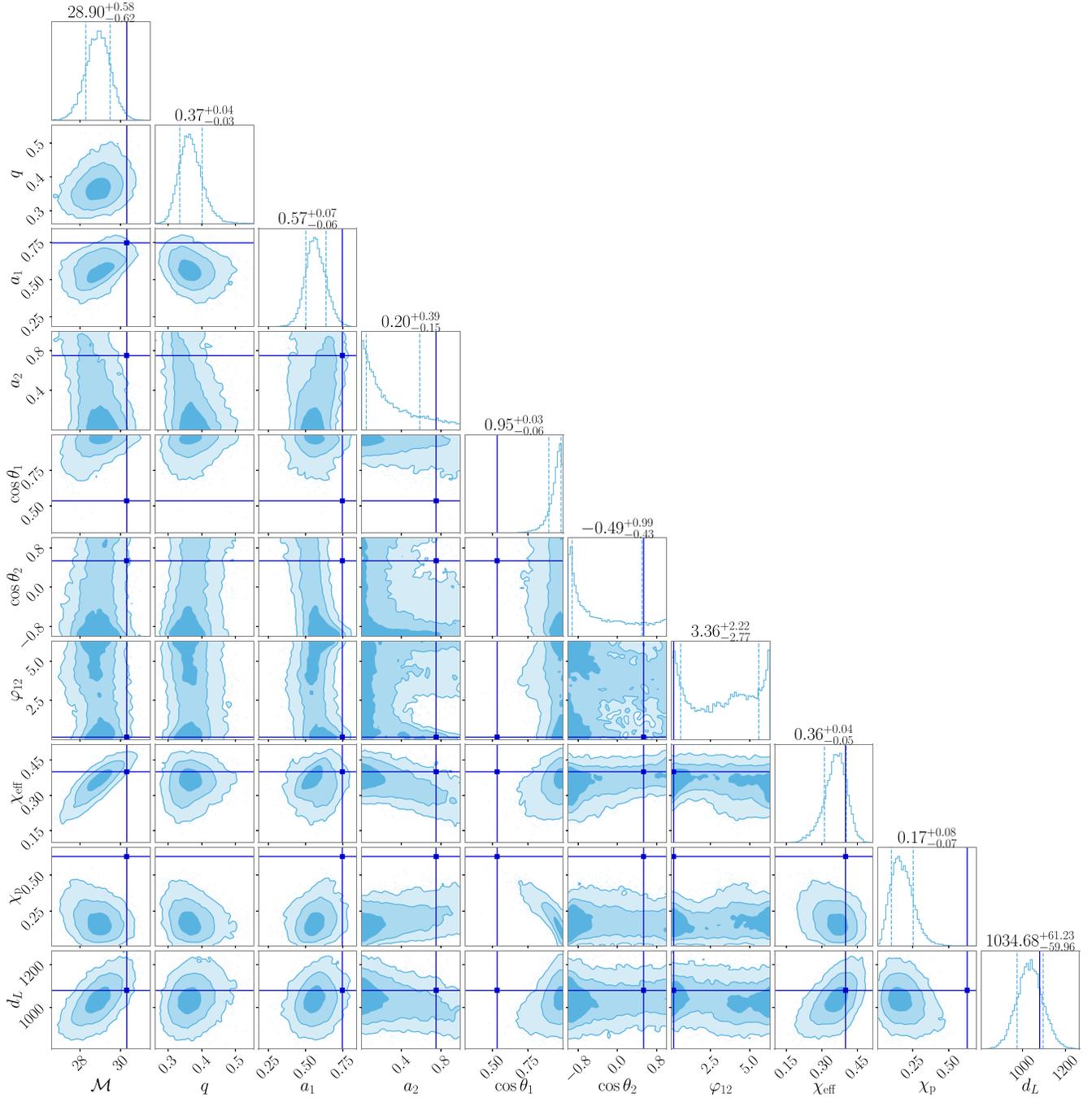

FIG. 20. Corner plot of the posteriors of injection B using IMRPhenomXPHM with SNR = 26.8. The plot includes the posteriors of the chirp mass, the mass ratio, the component spin magnitudes, the spin tilts, the spin azimuthal angle $\varphi_{12}$, $\chi_{\text{eff}}$, $\chi_{\text{p}}$, and the luminosity distance. The true parameter values are indicated by the vertical lines in dark blue.

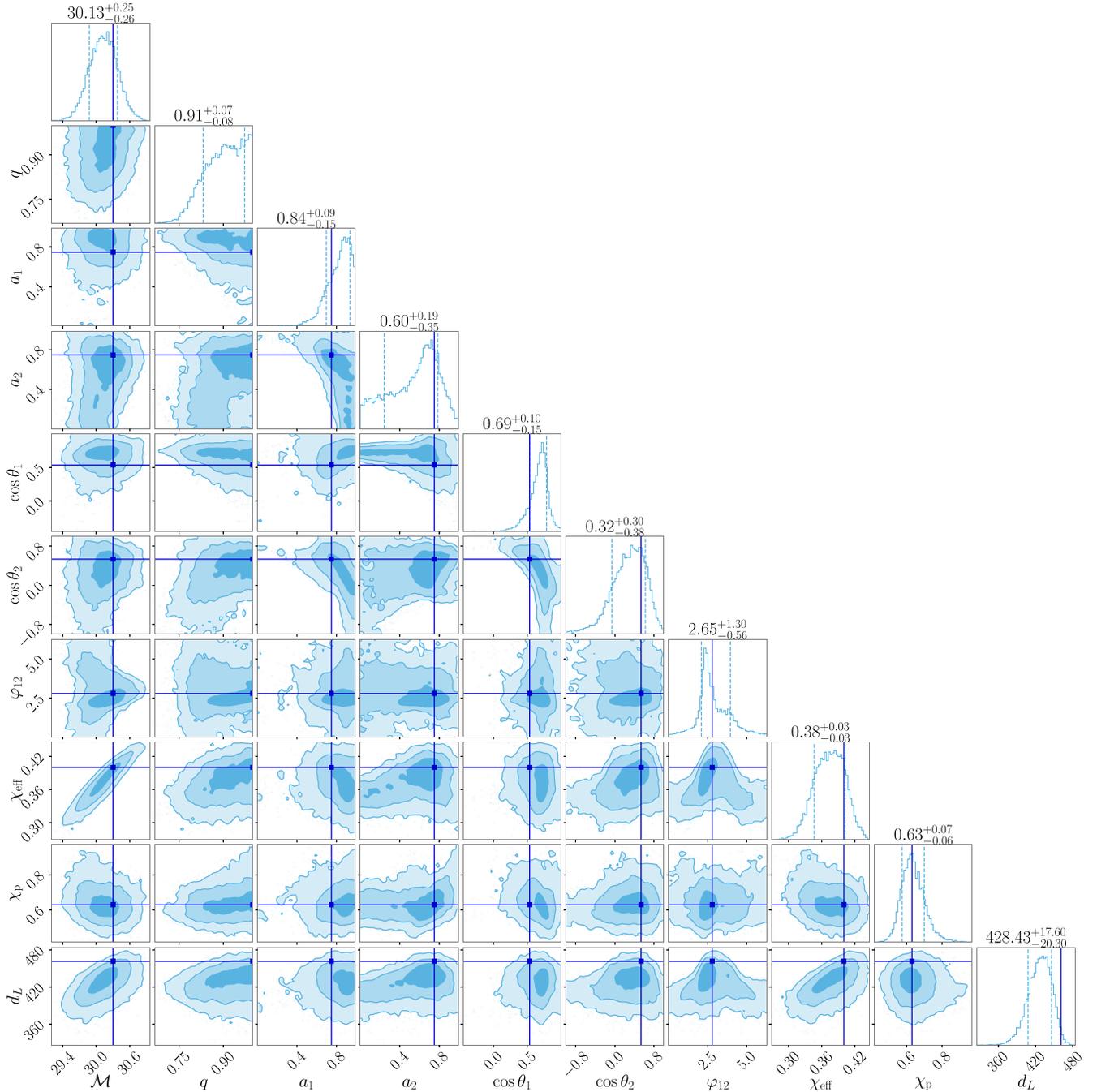

FIG. 21. Corner plot of the posteriors of injection A using IMRPhenomXO4a with SNR = 60. The plot includes the posteriors of the chirp mass, the mass ratio, the component spin magnitudes, the spin tilts, the spin azimuthal angle $\varphi_{12}$, $\chi_{\rm eff}$, $\chi_{\rm p}$, and the luminosity distance. The true parameter values are indicated by the vertical lines in dark blue.

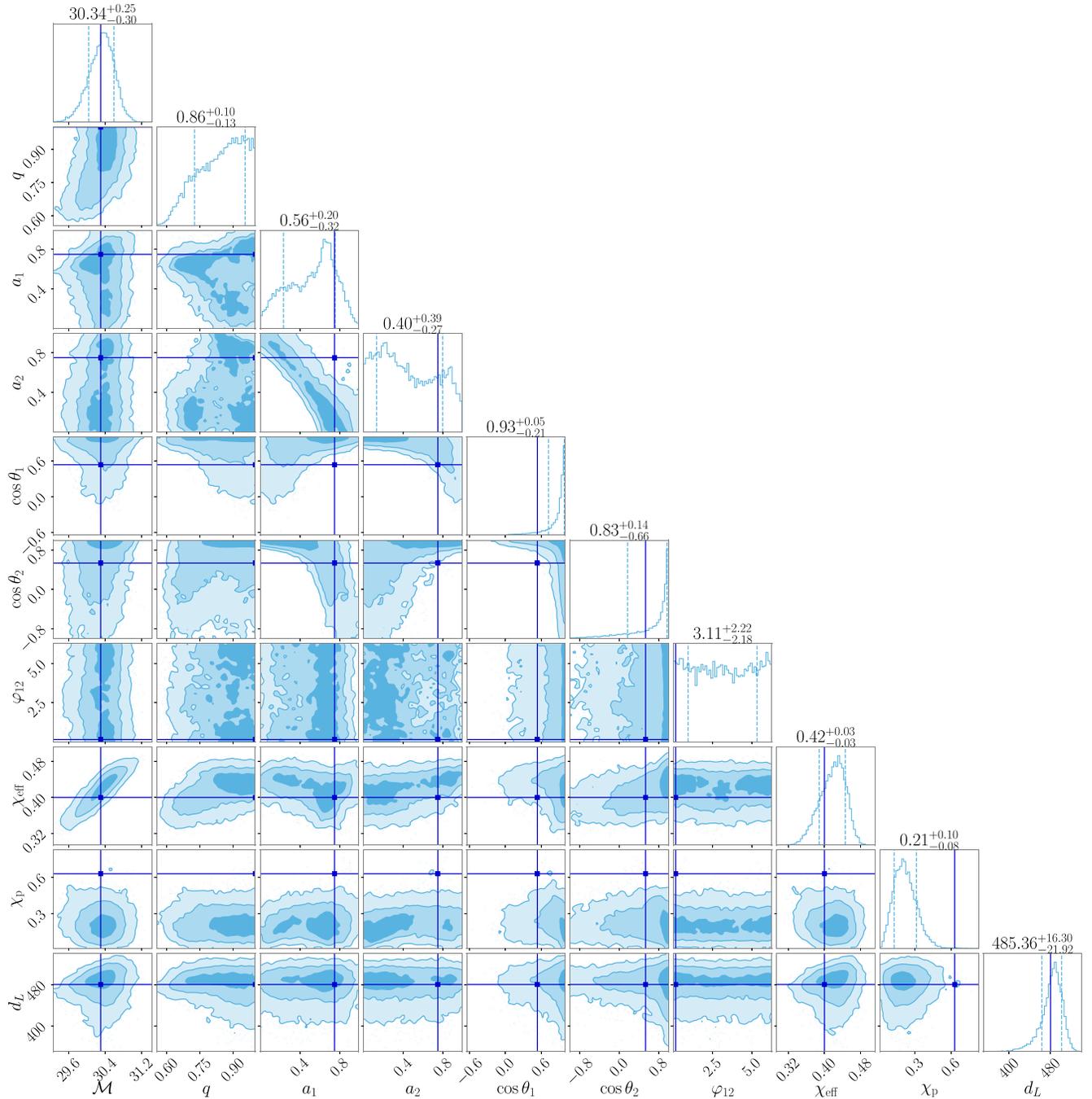

FIG. 22. Corner plot of the posteriors of injection B using IMRPhenomXO4a with SNR = 60. The plot includes the posteriors of the chirp mass, the mass ratio, the component spin magnitudes, the spin tilts, the spin azimuthal angle $\varphi_{12}$, $\chi_{\text{eff}}$, $\chi_{\text{p}}$, and the luminosity distance. The true parameter values are indicated by the vertical lines in dark blue.